\documentclass[namedreferences]{SolarPhysics}
\usepackage[optionalrh]{spr-sola-addons} 
\usepackage{graphicx}        
\usepackage{color}           
\usepackage{url}             
\def\al{&\!\!\!\!}

\newcommand{\aap}{    {\it Astron. Astrophys. }}

\newcommand{\apj}{    {\it Astrophys. J. }}
\newcommand{\apjl}{   {\it Astrophys. J. Lett. }}

\newcommand{\solphys}{{\it Solar Phys. }}

\begin{document}

\begin{article}

\begin{opening}
\title{ Slow Mode Oscillations and Damping of Hot Solar Coronal Loops \\{\it Solar Physics}}
\author{A.~ \surname{Abedini}$^{1}$\sep
        H.~\surname{Safari}$^{2}$\sep
        S.~\surname{ Nasiri}$^{2}$ }
\runningauthor{Abedini et al.}
 \runningtitle{Slow Mode
Oscillations of Coronal Loops ...}
   \institute{$^{1}$ Department of Physics, Institute for Advanced Studies in Basic Sciences,
 Zanjan, 45137-66731, I. R.  Iran
\\
                    $^{2}$ Department of Physics, University of Zanjan, University Blvd., 45371-38791, Zanjan, I. R.
                    Iran\\
                    email: \url{safari@znu.ac.ir}
             }
\begin{abstract}

The effect of temperature inhomogeneity on the periods, their
ratios (fundamental vs. first overtone), and the damping times of
the standing slow modes in gravitationally stratified solar
coronal loops are studied. The effects of optically thin
radiation, compressive viscosity, and thermal conduction are
considered. The linearized one-dimensional magnetohydrodynamic
(MHD) equations (under low-$\beta$ condition) were reduced to a
fourth--order ordinary differential equation for the perturbed
velocity. The numerical results indicate that the periods of
non-isothermal loops ({\it i.e.} temperature increases from the
loop base to apex) are smaller  compared to those of isothermal
loops. In the presence of radiation, viscosity, and thermal
conduction, an increase in the temperature gradient is followed
by a monotonic decrease in the periods (compared with the
isothermal case), while the period ratio turns out to be a
sensitive function of the gradient of the temperature and the
loop lengths. We verify that radiative dissipation is not a main
cooling mechanism of both isothermal and non-isothermal hot
coronal loops and has a small effect on the periods. Thermal
conduction and compressive viscosity are primary mechanisms in
the damping of slow modes of the hot coronal loops. The periods
and damping times in the presence of compressive viscosity and/or
thermal conduction dissipation are consistent with the observed
data in specific cases. By tuning the dissipation parameters, the
periods and the damping times could be made consistent with the
observations in more general cases.
\end{abstract}
\keywords{Sun, corona; Sun, magnetic fields; Sun, oscillations}
\end{opening}
\section{Introduction}

Through SOHO and TRACE observations, propagating slow
magnetoacoustic waves have been found in coronal plumes, loop
footpoints, above sunspots and even non-sunspot regions ({\it
e.g.}, Ofman {\it et al.}, 1997; DeForest and Gurman, 1998;
Berghmans and Clette, 1999; De Moortel {\it et al.}, 2000;
\'{O}Shea {\it et al.}, 2002; Brynildsen {\it et al.}, 2002;
Marsh {\it et al.}, 2003). Standing longitudinal slow waves with
strong damping and large Doppler-shift oscillations have been
detected in hot postflare loops recorded by SOHO/SUMER. In the
cooler loops, these waves were observed by the EUV imaging
spectrometer {\it Hinode}/EIS (Srivastava and Dwived, 2010).
These oscillations have a phase shift of about one-quarter period
between velocity and intensity. The periods and damping times of
the standing slow waves are in the ranges of 8.6--32.3 min and
3.1--42.3 min, respectively (Kliem {\it et al.}, 2002; Wang {\it
et al.}, 2002a, 2002b, 2003, 2005; Banerjee {\it et al.}, 2007;
Erd\'{e}lyi {\it et al.}, 2008). The effect of energy dissipation
on the slow waves through thermal conduction, compressive
viscosity, radiative cooling, and heating on the periods, period
ratio, and damping times has formed the focus of recent studies.
Ofman and Wang (2002), for instance, studied the oscillations and
damping of standing slow modes in the isothermal loops. They
found that thermal conduction is the dominant dissipation
mechanism. Taking into account the effects of thermal conduction
and compressive viscosity, De Moortel and Hood (2003)
investigated both propagating and standing slow magnetoacoustic
waves in homogeneous corona loops. They found a minimum damping
time that can be obtained by thermal conduction alone. However,
for stronger dissipation an additional mechanism such as
viscosity has to be added. Sigalotti {\it et al.} (2007) studied
the dissipation of standing slow modes in hot, isothermal loops
by integrating the effects of gravitational stratification,
thermal conduction, compressive viscosity, radiative cooling, and
heating in their study. They concluded that thermal conduction
and compressive viscosity are the main sources of the wave
damping. Pandey and Dwivedi (2006) have shown that separate
effects of thermal conduction and viscosity in isothermal loops
are not sufficient to explain the observed damping times of the
oscillations. Only with the combined effect of thermal conduction
and viscosity the results are consistent with the observations.
Moreover, Taroyan {\it et al.} (2005) investigated the effect of
temperature inhomogeneity on the dissipation of standing slow
waves through considering thermal conduction and optically thin
radiative losses. They found that the damping time of the
isothermal loops is proportional to the wave period and the
oscillations are rapidly damped mainly by thermal conduction.

The ratio between the period of the fundamental mode ($p_1$) and
the first overtone ($p_2$), $p_1/(2p_2)$, for both the fast kink
mode and the slow longitudinal mode is a useful seismological
indicator. Departure of this ratio from unity is a consequence of
the density, pressure, temperature, magnetic field (both radial
and longitudinal structures), heating functions and the
dissipation mechanisms ( D\'{i}az {\it et al.}, 2006; Dymova and
Ruderman, 2006; Donnelly {\it et al.}, 2006; Erd\'{e}lyi and
Verth, 2007; Safari {\it et al.}, 2007; Verth {\it et al.}, 2007;
Ruderman {\it et al.}, 2008; McEwan {\it et al.}, 2008; Andries
{\it et al.}, 2009; Fathalian and Safari, 2010). The results of
the numerical modeling of the oscillations allow a comparison
between the model and the observational results (McEwan {\it et
al.}, 2006; Abedini and Safari, 2011). Macnamara and Roberts
(2010) studied the effects of thermal conduction and compressive
viscosity on the slow mode of the isothermal loops, and concluded
that the effect of thermal conduction on the period ratio is
negligible.

In line with the above-mentioned studies, in this study, we will
investigate the dissipation of the standing slow MHD modes in the
hot coronal loops taking into account gravitational
stratification, inhomogeneity of temperature, thermal conduction,
compressive viscosity, heating, and optically thin radiative
losses.

For such a purpose, the linearized one dimensional
magnetohydrodynamic (MHD) equations are reduced to a fourth--order
differential equation for the perturbed velocity. The paper is
organized as follows. In Section 2, we present a brief
description of the models and equations. In Section 3, the
differential equation is solved numerically based on the boundary
value problem solver (finite difference code) for different
cases. Finally, conclusions are drawn in Section 4.

\section{Description of the Model and Equations}
Coronal loops are modeled as a magnetic flux tube of length $2L$
with a shape of half-circle and symmetrical around the apex
($z=0$), with a constant magnetic field along the loop length.
The footpoints are fixed in the chromosphere. The equilibrium
plasma flow is ignored, ${\bf v}_0=0$. The projection of the
gravity acceleration on the tangent to the loop (half-circle) is
equal to $g \sin \pi z/2L$. The equilibrium temperature is
considered as a function of $z$,
\begin{eqnarray}
\label{equ1} \al \al T_0(z)=T_0(L)\left[1+(\lambda-1)\cos\frac{\pi
z}{2L}\right],~ -L\leq z\leq L,~~\lambda=\frac{T_0(0)}{T_0(L)},
\end{eqnarray}
where $T_0(0)$ and $T_0(L)$ are the temperatures at the apex and
at the base, respectively. Under the ideal gas law as well as in
the low-$\beta$ plasma, the equilibrium density and pressure are
given by
\begin{eqnarray}
\label{equ2}
\al\al \rho_0(z)=\rho_0(L)\left[1+(\lambda-1)\cos \frac{\pi z}{2L}\right]^{-(\frac{N}{\alpha_T}+1)},\nonumber\\
\al\al p_0(z)=p_0(L)\left[1+(\lambda-1)\cos\frac{\pi
z}{2L}\right]^{-\frac{N}{\alpha_T}},
~\alpha_T=\frac{\pi(\lambda-1)}{2},
\end{eqnarray}
where $ N=L/\Lambda_0(L)$ and the pressure scale height is
defined by $\Lambda_0={p_0}/{\rho_0g}$. The energy terms, the
constant heating rate ($H=H_0$), the conductive heating ($E_{\rm
C}$), the radiative cooling ($E_{\rm R}$), and the compressive
viscous heating ($E_\eta$) all are given by
\begin{eqnarray}
\label{equ3}
 E_{\rm C}=\frac{\partial } {\partial z}( k_{||}\frac{\partial T}{\partial z} ),~
 E_{\rm R}=\chi{\rho}^{2} T^{\alpha},~E_\eta=\frac{4}{3} \eta(\frac{\partial v}{\partial z})^2,
\end{eqnarray}
in which $k_{||}=10^{-11}T^{5/2}(\rm W \rm m^{-1}\rm K^{-1})$ and
$\eta=10^{-17}T^{5/2}(\rm kg\ \rm m^{-1}\rm s^{-1})$ are the
coefficient of thermal conduction along the magnetic field and
the coefficient of compressive viscosity, respectively
(Braginskii, 1965; Hildner, 1974; Priest, 1982). The parameters
$\alpha$ and $\chi$ are $\alpha = -1$ and $\chi = 10^{24} {\rm
W}{\rm m}^{3}{\rm kg}^{-2}{\rm K}$, respectively, for the
temperature range of $0.8- 10$ MK (Sigalotti {\it et al.}, 2007).
In the equilibrium state, we get $ H_0+E_{\rm 0C}+E_{0
\eta}=E_{\rm 0R}$. For the convenience of our analysis, the
dimensionless parameters are defined as
\begin{eqnarray}
\label{equ4} \al\al \bar{t}=t/\tau_{\rm
s},~\bar{z}=\frac{z}{L},~\bar{v}=\frac{v}{c_{\rm s}(L)},
\bar{T}=\frac{T}{T_0(L)},~\bar{p}=\frac{p}{p_0(L)},~\bar{\rho}=\frac{\rho}{\rho_0(L)},\nonumber\\
\al\al\bar{\eta}=\frac{\eta}{\eta_0(L)},~
\bar{k_{||}}=\frac{k_{||}}{k_{0||}(L)}, r=\frac{(\gamma-1)\tau\chi\rho_0^2(L)T_0^{-1}(L)} {\gamma p_0(L)},\nonumber\\
 \al\al \epsilon=\frac{1}{\rm Re}=\frac{\eta_0(L)\tau_{\rm s}} {\rho_0(L)L^2},~~
 d=\frac{(\gamma-1)k_{0||}(L)T_0(L)\rho_0(L)} {\gamma^2 p_0^2(L)\tau_{\rm s}},
\end{eqnarray}
where $c_{\rm s}^2(L)=\gamma P_0(L)/\rho_0(L)$ is the adiabatic
sound speed, $\tau_{\rm s}={L}/{c_{\rm s}}$ is the sound crossing
time from the loop footpoints to the apex, and Re is the Reynolds
number. The dimensionless parameters $d$, $r$, and $\epsilon$ are
called the thermal ratio, radiation ratio, and the compressive
viscosity ratio, respectively ({\it e.g.}, De Moortel and Hood,
2003, 2004). In hot coronal plasma conditions, the values of $r$,
$\epsilon$, $d$ and $\tau_{\rm s}$ are tabulated in Table 1.

\begin{center}
\begin{table}[ht]
\caption{%
Dimensionless ratios and the sound crossing time as functions of
the loop length ($L$) and the apex temperature $(T_0(L))$.
}\centering
\begin{tabular}{c c c c c c}
\hline
$T_0(L)$ [MK]& $L$ [Mm] & $r\times 10^{3}$ & $\epsilon$ & $d$ & $\tau_{\rm s} [{\rm min}]$ \\
\hline
 6 & 20--220 & 0.05--0.53 & 0.02--0.21 &0.54--6.0& 0.88--9.64 \\
\hline
 10 & 20--220 &0.01--0.11 & 0.08--0.83 &2.18--24.0 & 0.69--7.65 \\
 \hline
\end{tabular} \label{jadval1}
\end{table}
\end{center}

In the low-$\beta$ condition, the linearized non-ideal MHD
equations are
\begin{eqnarray}
\al\al\frac{\partial \rho_1} {\partial{t}}+\rho_0\frac{\partial
v_1}{\partial z}+\frac{\partial\rho_0}{\partial z}v_1=0,
\label{equ5.1}\\
 \al\al
\rho_0\frac{\partial{ v_1}} {\partial{t}} =-\frac{\partial p_1}
{\gamma\partial z}
+\frac{ N \sin(\frac{\pi z}{2})\rho_1}{\gamma}+\eta_0\varepsilon\frac{4}{3}\frac{\partial^2 v_1 } {\partial z^2},\\
\al\al
\frac{p_1}{p_0}=\frac{\rho_1}{\rho_0}+\frac{T_1}{T_0},\nonumber\\
\al\al\frac{\partial{p_1}}{\partial{t}}+{v_1}\frac{\partial
p_0}{\partial z} +\gamma p_0\frac{\partial v_1}{\partial z}=
\nonumber\\\al\al~~~~~~\gamma \left( -2r E_{\rm
0R}\frac{\rho_1}{\rho_0}+r E_{\rm 0R}\frac{T_1}{T_0} + d
\frac{\partial}{\partial z}(k_{0||} \frac{\partial
T_{1}}{\partial z} + k_{1||}\frac{\partial T_0}{\partial
z})\right).\label{equ5.2}
\end{eqnarray}
As we note here, all the quantities in Equations
(\ref{equ5.1})--(\ref{equ5.2}) are dimensionless (bars on
dimensionless quantities have been dropped). The perturbed
quantities are assumed to be of the form $f_1(z,t)=f_1(z)\exp(\nu
t)$. After some algebra, Equations (\ref{equ5.1})--(\ref{equ5.2})
reduce to a fourth-order differential equation as
\begin{eqnarray}
\label{equ6} \al\al {\cal A}(z,r,\epsilon,d,\nu)\frac{d^4v_1
}{dz^4}+{\cal B}(z,r,\epsilon,d,\nu)\frac{d^3v_1 }{dz^3}
 +{\cal C}(z,r,\epsilon,d,\nu)\frac{d^2v_1 }{dz^2}\nonumber\\
\al\al +{\cal D}(z,r,\epsilon,d,\nu)\frac{dv_1 }{dz}+{\cal
E}(z,r,\epsilon,d,\nu)=0.
\end{eqnarray}
Coefficients ${\cal A}$, ${\cal B}$, ${\cal C}$, ${\cal D}$, and
${\cal E}$ are given in the Appendix.

The boundary conditions are
\begin{itemize}
\item The footpoints ($z=\pm 1)$ are expected to be the nodes, {\it i.e.}, $v|_{\rm footpoints}=0$.
\item Near the apex, $z\to0$, all the coefficients in Equations (\ref{equ5.1})--(\ref{equ5.2}) are constants and the solutions are expected to be $v(z,t)\sim\exp[-i(\omega t-kz)]$.
\end{itemize}
In the remainder of this paper, Equation (\ref{equ6}) is solved
for five various cases by imposing the appropriate boundary
conditions.

\section{Solutions to Equation (8)}
In all of the following cases, the loops are considered to be
gravitationally stratified.

\subsection{Case 1: $d=r=\epsilon=0, ~N\neq0$}
\begin{figure} 
\centerline{\includegraphics[width=1\textwidth,clip=]{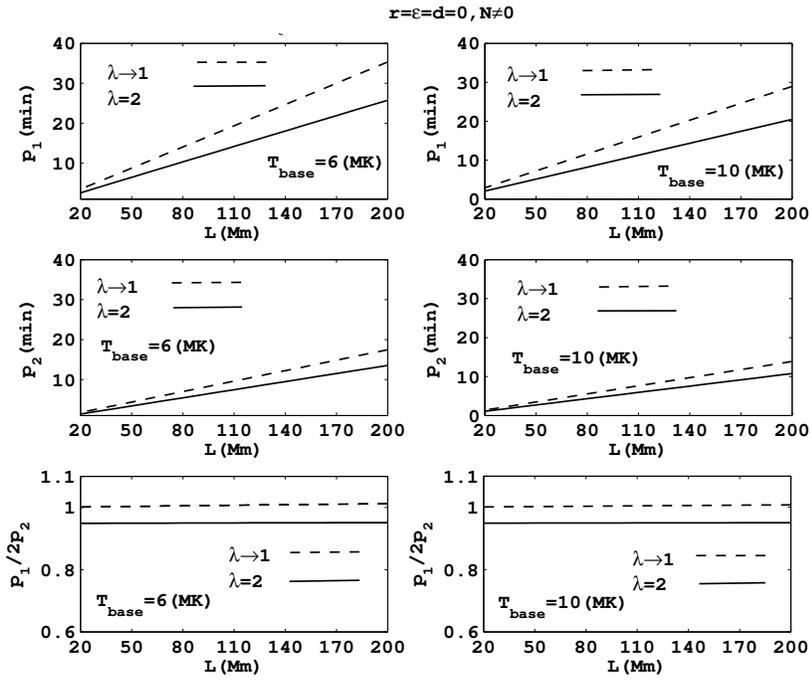}}
\caption[]{%
The periods of the fundamental mode $p_1$ (top row) and of the
first overtone $p_2$ (middle row), and their ratios (bottom row)
for case 1 (see text) are plotted versus $L$ in oscillating
stratified loops with different values of $\lambda$ and $T_{\rm
base}$. } \label{fig1}
\end{figure}

In the absence of all dissipation terms, Equation (\ref{equ6})
reduces to a second order ordinary differential equation. The
resultant equation is solved numerically. In Figure \ref{fig1},
the fundamental and the first overtone modes and their ratios,
$p_1/(2p_2)$, are plotted versus $L$. As shown in the figure, in
the case of adiabatic and isothermal loops ($\lambda\rightarrow
1$), the fundamental and first overtone periods are in the ranges
of 3--32 and 1--18 min, respectively. The results are well
consistent with the previous studies (Curdt {\it et al.}, 2002;
McEwan {\it et al.}, 2006; Sigalotti {\it et al.}, 2007; Abedini
and Safari, 2011).

We see that, by increasing $\lambda$ from 1 to 2 the periods and
their ratio decrease; yet, the changes in the fundamental periods
are larger than those of the overtone periods. For typical loop
lengths of 20--200 Mm, after adding the temperature gradient
($\lambda =2$) the changes in the periods is found to be about a
few minutes ($\le$10).

\subsection{Case 2: $d=\epsilon=0, ~r\neq0, ~N\neq0$}
\begin{figure} 
\centerline{\includegraphics[width=1\textwidth,clip=]{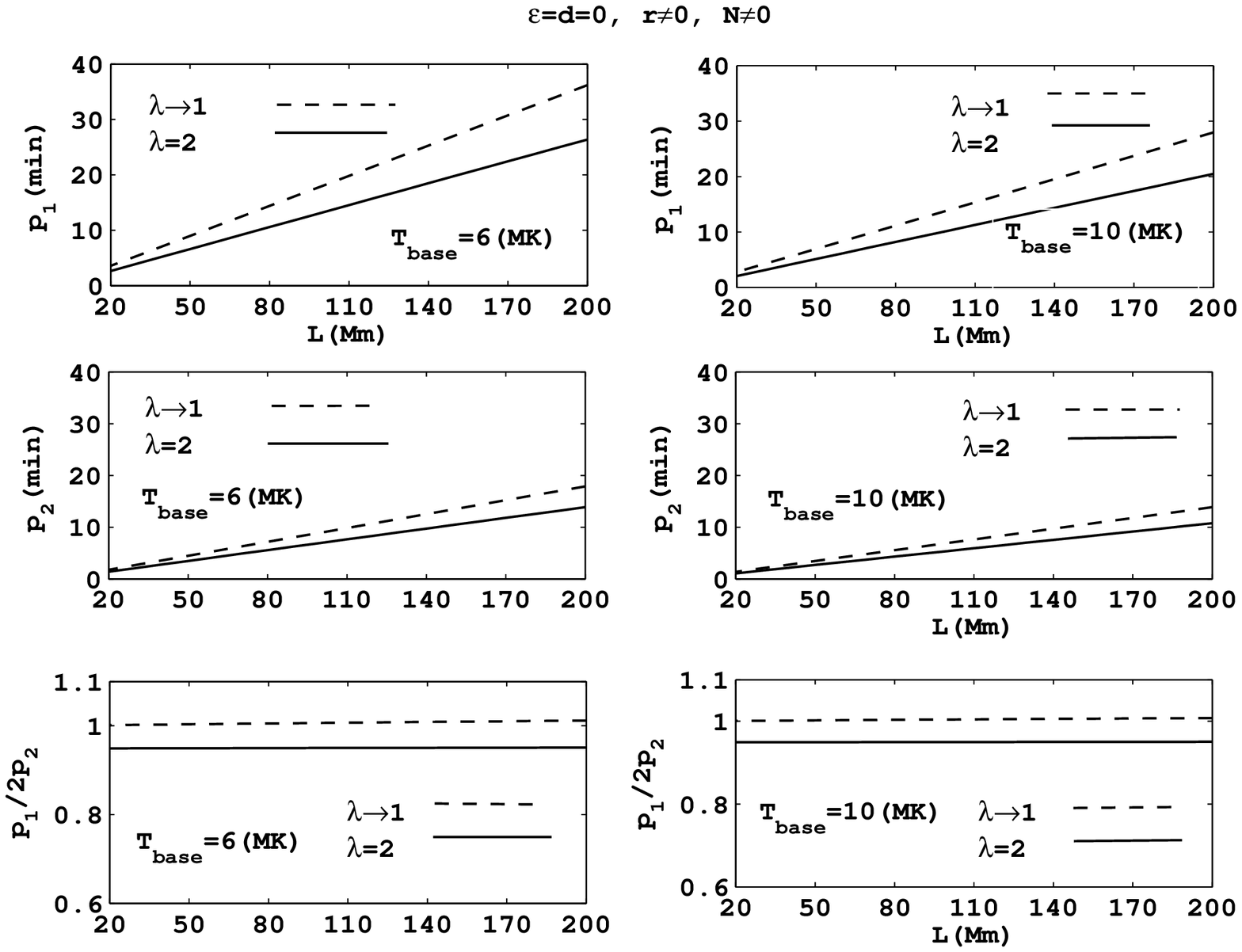}}
\caption[]{%
The periods of the fundamental mode $p_1$ (top row) and of the
first overtone $p_2$(middle row), and their ratios (bottom row)
for case 2 (see text) are plotted versus $L$ in oscillating loops
in the presence of radiation and gravity for different value of
$\lambda$ and $T_{\rm base}$. } \label{fig2}
\end{figure}

\begin{figure} 
\centerline{\includegraphics[width=1\textwidth,clip=]{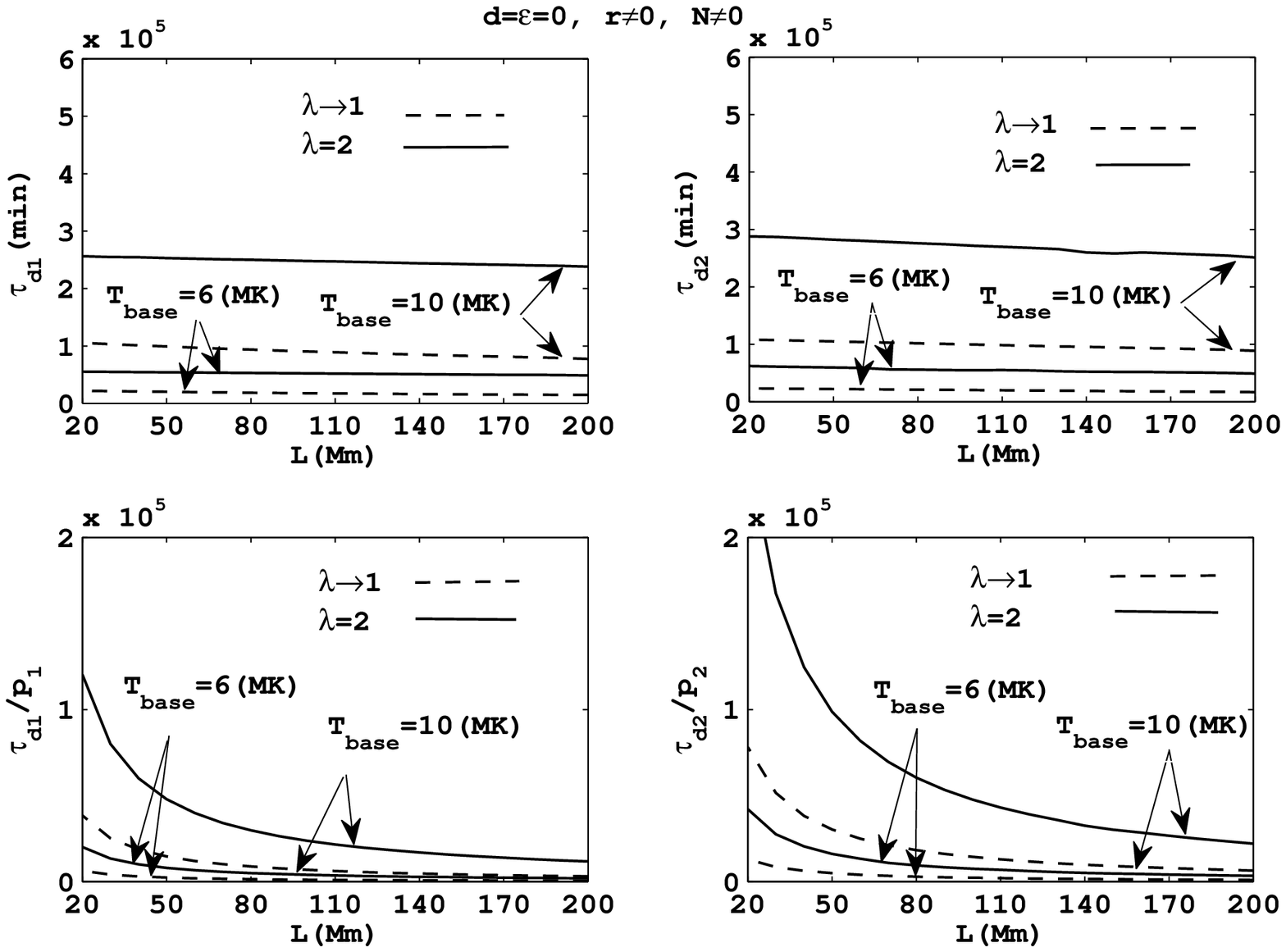}}
\caption[]{%
The damping times of the fundamental mode ($\tau_{\rm d1}$) and of
the first overtone ($\tau_{\rm d2}$) (top row), and the damping
times per period ($\tau_{\rm d1}/p_1, \tau_{\rm d2}/p_2$) (bottom
row) for case 2 (see text) are plotted versus $L$ in oscillating
loops in the presence of radiation and gravity for different
value of $\lambda$ and $T_{\rm base}$. } \label{fig3}
\end{figure}

In the presence of optically thin radiation ({\it i.e.},
$\epsilon=d=0, r\neq0$), similar to the previous case, we have a
second order differential equation. The numerical solutions to
the resultant equation are plotted in Figures \ref{fig2} and
\ref{fig3}. By comparing Figures \ref{fig1} and \ref{fig2}, we
find that the fundamental and the first overtone modes have
changed about one percent. Therefore we conclude that the
radiation does not change the periods and their ratios
significantly. In Figure \ref{fig3}, the damping time and damping
quality (damping per period, $\tau_{\rm d}/p={\rm
Re}(\omega)/2\pi {\rm Im}(\omega)$) are plotted versus $L$ for
different $T_{\rm base}$ and $\lambda$. As shown in the figures,
the damping time is of the order of $10^6$ min and the damping
quality is of the order of  $10^5$. This case lies in the weak
damping regime ($\tau_{\rm d}/p\geq 2$). By increasing $\lambda$
from 1 to 2, the damping time and damping quality increase. We
arrive at the conclusion that, in the hot coronal loop, radiation
is not a primary mechanism for the damping of the slow modes.

\subsection{Case 3: $r=d=0,~\epsilon\neq0, ~N\neq0$}
\begin{figure} 
\centerline{\includegraphics[width=1\textwidth,clip=]{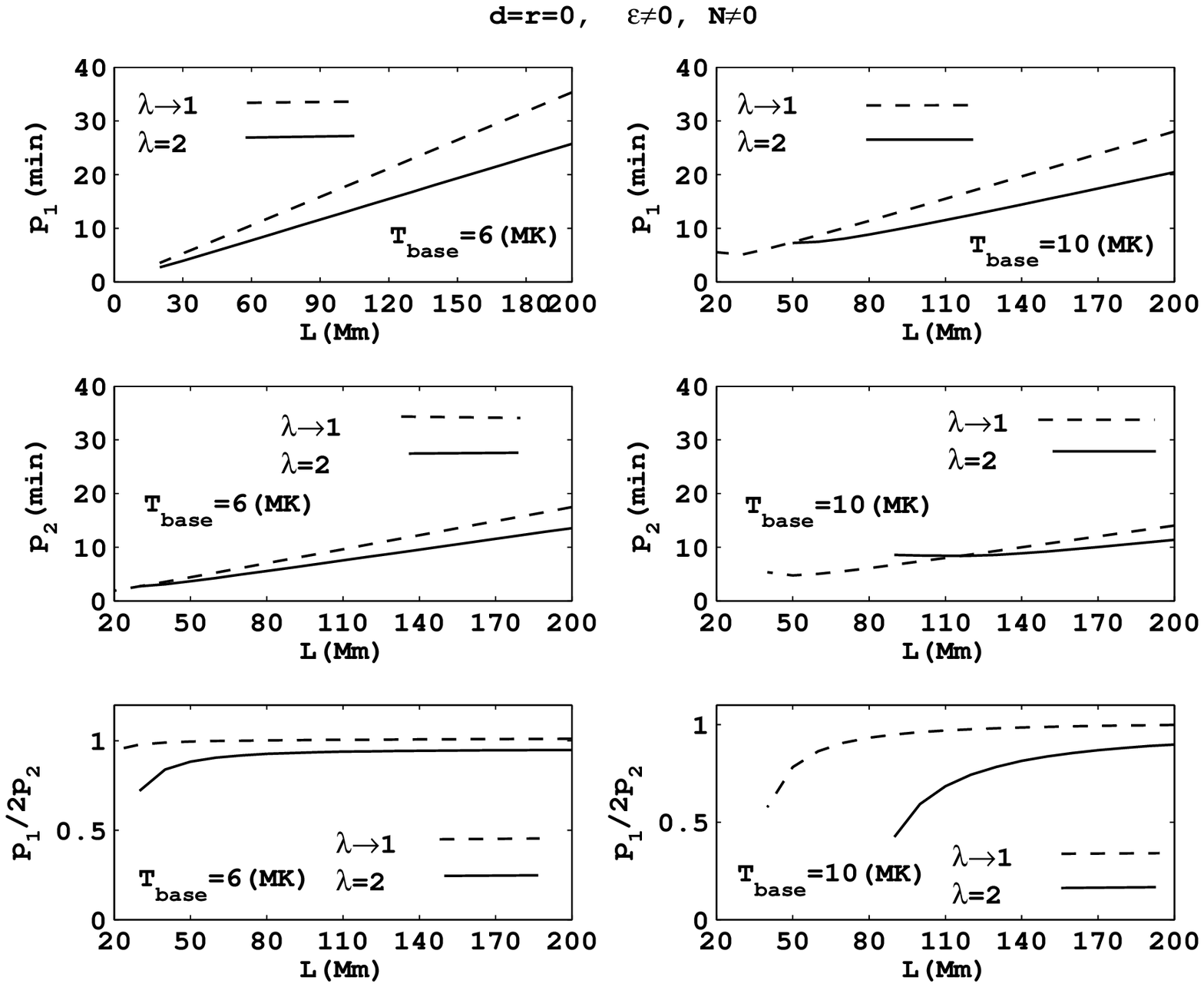}}
\caption[]{%
The periods of the fundamental mode $p_1$ (top row) and of the
first overtone $p_2$(middle row), and their ratios (bottom row)
for case 3 (see text) are plotted versus $L$ in oscillating loops
in the presence of viscosity and gravity for different value of
$\lambda$ and $T_{\rm base}$. } \label{fig4}
\end{figure}

\begin{figure} 
\centerline{\includegraphics[width=1\textwidth,clip=]{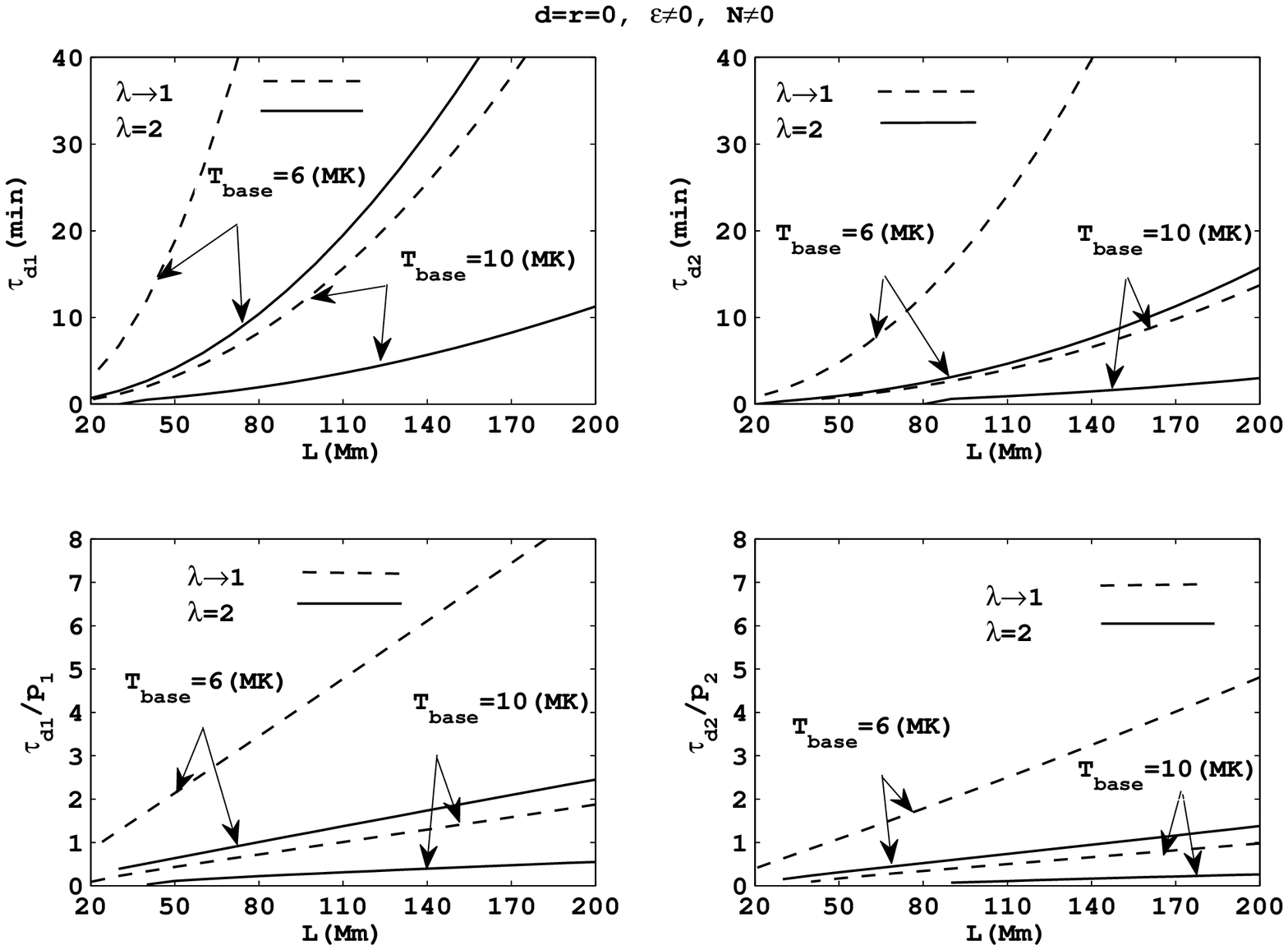}}
\caption[]{%
The damping times of the fundamental mode ($\tau_{\rm d1}$) and of
the first overtone ($\tau_{\rm d2}$) (top row), and the damping
times per period ($\tau_{\rm d1}/p_1, \tau_{\rm d2}/p_2$) (bottom
row) for case 3 (see text) are plotted versus $L$ in oscillating
loop in the presence of viscosity and gravity for different value
of $\lambda$ and $T_{\rm base}$. } \label{fig5}
\end{figure}

Next, the effects of compressive viscosity on the oscillations
and damping of the loops are investigated. The numerical
solutions are shown in Figures \ref{fig4} and \ref{fig5}. By
comparing Figures \ref{fig1} and \ref{fig4} we see the following:
Compressive viscosity introduces a cutoff frequency on the
oscillations as a function of both inhomogeneity parameter
$\lambda$ and loop length $L$ (see, {\it e.g.}, Sigalotti {\it et
al.}, 2007). The effect of compressive viscosity on the periods
for a typical loop length of 20--200 Mm is a few percent ($\le
2\%$) increase compared to case 1. By increasing inhomogeneity in
the temperature ($\lambda$) and decreasing the loop length, the
ratio ($p_1/(2p_2)$) deviates from unity. At higher temperatures
($\geq 6$ MK) and non-isothermal loops, viscous dissipation is
one of the primary damping mechanisms. Indeed, viscosity is one
of the main factors for the strong damping in the short and
non-isothermal loops and is consistent with the observational
data.

In the presence of the temperature gradient, the damping of the
first overtone is stronger than that of the fundamental mode.
Unlike radiation (case 2), if $\lambda$ is increased from 1 to 2
in the presence of viscosity, the damping time and damping
quality decrease. We conclude that, for non-isothermal loops in
the presence of viscosity and gravity, the computed damping times
are consistent with the SUMER observations (Wang {\it et al.},
2003, 2005).

\subsection{Case 4: $d\neq0,r=\epsilon=0, ~N\neq0$}
In the presence of thermal conduction, the periods decrease
slightly if the temperature increases (Figure 6). However, by
increasing inhomogeneity in the temperature ($\lambda$ from 1 to
2), the periods and their ratio decrease considerably. In other
words, the decrease in the period is more sensitive to $\lambda$
than the temperature itself. The effect of thermal conduction on
the periods compared to the stratified isothermal loops (case 1,
$\lambda\rightarrow 1$) is about 18 min. For $\lambda =2$ the
increase in the periods is about 8 min. The deviation in the
period ratio from unity increases with increasing $L$ and
$\lambda$ (for $T_{\rm base}=6 $ MK). The deviation of the
damping quality from unity in isothermal loops decreases with
increasing $L$ for short loops (Figure 7). However, if the loop
length is further increased, the damping quality of
non-isothermal loops  becomes larger than that of the isothermal
case.

There is a negative relationship between the damping time and
temperature in the case of non-isothermal loops: An increase in
the temperature brings about a decrease in the damping time. On
the other hand, there is a positive relationship between the
damping time and the loop length. We see that, in the presence of
thermal conduction, the damping of both isothermal and
non-isothermal long loops are in the strong regime ($\tau_{\rm
d}/p\leq 1$).

\subsection{Case 5: $d\neq0,\epsilon\neq0, ~r\neq0, ~N\neq0$}
In the most general case which includes heat conduction,
compressive viscosity, and optically thin radiation, Equation
(\ref{equ6}) must be solved. The numerical results are presented
in Figures \ref{fig8} and \ref{fig9}. In Figure \ref{fig8},
periods and their ratios are plotted versus $L$. Compared with
the case of the adiabatic loops (Figure 1), the periods increase
noticeably. Periods with $T_{\rm base}=6$ and 10 MK are 2--50 min
and 1--45 min, respectively (for isothermal loops). By increasing
$\lambda$ from 1 to 2, the periods decrease about 20 min.

In the presence of all dissipation mechanisms and temperature
gradient, the damping time and damping quality are consistent
with the observed values (Banerjee {\it et al.}, 2007; Kliem {\it
et al.}, 2002; Wang {\it et al.}, 2002a, 2002b, 2003, 2005).

\section{Summary}
The coronal loop has been modeled as a half-circle magnetic flux
tube with a uniform magnetic field along the loop axis. The
effect of the inhomogeneity of the temperature on the periods,
the period ratio, and damping times of standing slow modes in the
gravitationally stratified coronal loops was investigated in this
study. The effects of optically thin radiation, compressive
viscosity, and thermal conduction were considered. The linearized
one-dimensional MHD equations were reduced to a fourth-order
ordinary differential equation for the perturbed velocity. Under
the solar coronal conditions (low-$\beta$ plasma) the resultant
equation was solved numerically. Moreover, the oscillations and
the damping of various flux tube models in the presence of
dissipation mechanisms were compared with the oscillations and
damping of adiabatic and isothermal loops. The following shows
the main results:

\begin{itemize}
\item In a gravitationally stratified loop, the periods
and their ratio (fundamental vs. first overtone) are
sensitive functions of the temperature inhomogeneity.
\item In hot corona loops ($T\ge6$ MK), the effect of
optically thin radiation on the periods and damping
of the oscillations is negligible for both isothermal
and non-isothermal cases.
\item In the presence of compressive viscosity, the damping
times significantly change but viscous dissipation is more
effective in shorter and hotter loops. Compressive viscosity
introduces a cutoff frequency on the oscillations.

In the case of non-isothermal and hotter loops the computed
damping times agree with the observed data.
\item Thermal conduction changes the periods more significantly
than other dissipation mechanisms, and the damping of oscillations
is in the weak regime ($\tau_{\rm d}/p\ge2$). We noted that thermal
conduction is a function of the temperature gradient and depends on
higher derivatives, and this is the reason for bringing more changes
in the periods in addition to the changes due to inhomogeneity in the temperature.
\item In the most general case where optically thin radiation,
compressive viscosity, and thermal conduction are all considered,
the periods are positively correlated with the loop length and
negatively correlated with the temperature. The damping time and
damping quality are complex functions of both loop length $L$ and
inhomogeneity parameter $\lambda$. We also found that in the
isothermal gravitationally stratified loops, thermal conduction
and compressive viscosity are needed to reproduce the observed
periods and damping times. In the presence of the temperature
gradient in the gravitationally stratified non-isothermal loops,
the periods and the damping times resulted by compressive viscosity
and/or thermal conduction dissipation, are consistent with the observed
data in special cases. However, by tuning the dissipation parameters, the
periods and the damping times could be made in good agreement with the
observations in more general cases.
\end{itemize}

\begin{acks}
The authors thank the anonymous referee for very helpful comments
and suggestions.
\end{acks}

\section*{Appendix}
Coefficient ${\cal A}$, ${\cal B}$, ${\cal C}$, ${\cal D}$, and ${\cal E}$ in Equation (8) are:\\
$T_0(z)=[1+(\lambda-1)\cos(\frac{\pi z}{2}],~
\rho_0(z)=T_0(z)^{-(\frac{N}{\alpha_T}+1)},~
p_0(z)=T_0(z)^{-\frac{N}{\alpha_T}},\\~g=\sin(\frac{\pi z}{2}),
~ k_{0||}=T_0^{5/2}(z),~\eta_0(z)=T_0^{5/2}(z),~\\
 f(z)=\gamma d k_{0||}[-\frac{T_0p_0''}{p_0^2}+\frac{2T_0p_0'^2}{p_0^3}
-\frac{7T_0'p_0'}{p_0^2}+\frac{35T_0'^2}{4p_0T_0}
+ \frac{7T_0''}{2p_0}] +r\gamma\rho_0 T_0^{-2} \\
{\cal A}(r,\epsilon,d,z)= \gamma d
k_{0||}[\frac{4\gamma\eta_0\epsilon \nu}{3\rho_0}
+T_0]\\
{\cal B}(r,\epsilon,d,z)= \gamma d
k_{0||}[(\frac{19T_0'}{2p_0}-\frac{2T_0p_0'}{p_0^2}-\frac{
\rho_0'}{\rho_0^2})\frac{4\gamma\eta_0\epsilon}{3} \nu
+\frac{4\gamma\eta_0'\epsilon}{\rho_0}\nu]
\\
+\gamma d k_{0||}[- N g
 +\frac{T_0\rho_0'}{\rho_0}
+\frac{21T_0'}{2}
-(T_0 +\frac{4\gamma\eta_0\epsilon}{3\rho_0} \nu)\frac{f(z)'}{(f(z)-\nu)}]\\
{\cal C}(r,\epsilon,d,z)= +\gamma d
k_{0||}[(\frac{35T_0'^2}{2T_0p_0}
+\frac{7T_0''}{p_0}-\frac{14T_0'p_0'}{p_0^2}-\frac{2T_0p_0''}{p_0^2}+\frac{4T_0p_0'^2}{p_0^3})\gamma\eta_0\epsilon\frac{4}{3}
\nu]
\\
 +\gamma d k_{0||}(\frac{19T_0'}{2p_0}-\frac{2T_0p_0'}{p_0^2}-\frac{\rho_0'}{\rho_0^2})[\gamma\eta_0'\epsilon\frac{8}{3}\nu - N g\rho_0 ]\\
+ \gamma d k_{0||}[\frac{3T_0\rho_0''}{\rho_0}
 +\frac{105T_0'^2}{4T_0}
+ \frac{21T_0''}{2}
 +\frac{21T_0'\rho_0'}{2\rho_0}-\frac{3T_0\rho_0'^2}{\rho_0^2} - 2N g' - \frac{3N g\rho_0'}{\rho_0}-\gamma\nu^2
]
\\
+\gamma d k_{0||}[N g-7T_0' -\frac{T_0\rho_0'}{\rho_0} +(-\frac{7T_0'}{p_0}+\frac{2T_0p_0'}{p_0^2})\gamma\eta_0\epsilon\frac{4}{3} \nu-\gamma\eta_0'\epsilon\frac{8}{3\rho_0}\nu )]\frac{f(z)'}{(f(z)-\nu)}\\
+3r\gamma \rho_0^2 T_0^{-1}
+(f(z)-\nu)\frac{4\gamma\eta_0\epsilon}{3}\nu -\gamma p_0 \nu
\\
{\cal D}(r,\epsilon,d,z)= +\gamma d
k_{0||}[\frac{35T_0'^2}{2T_0p_0}
+\frac{7T_0''}{p_0}-\frac{14T_0'p_0'}{p_0^2}-\frac{2T_0p_0''}{p_0^2}+\frac{4T_0p_0'^2}{p_0^3})(\gamma\eta_0'\epsilon\frac{4}{3}\nu - N g\rho_0) ]\\
+\gamma d
k_{0||}[\frac{21T_0'\rho_0''}{\rho_0}+\frac{3T_0\rho_0'''}{\rho_0}-\frac{9T_0\rho_0'\rho_0''}{\rho_0^2}
 -\frac{21T_0'\rho_0'^2}{\rho_0^2}+\frac{6T_0\rho_0'^3}{\rho_0^3}+ \frac{21\rho_0'T_0''}{2\rho_0}\\
+\frac{105T_0'T_0''}{4T_0} +\frac{105T_0'^2\rho_0'}{4\rho_0T_0} +
\frac{7T_0'''}{2} +\frac{ 105T_0'^3}{8T_0^2}
] \\
+\gamma d k_{0||}[\frac{4\gamma\eta_0'''\epsilon}{3\rho_0}\nu
- N g''-\frac{ 4N g'\rho_0'}{\rho_0}- \frac{3N g\rho_0''}{\rho_0}-\frac{2\gamma\rho_0'\nu^2}{\rho_0} ]\\
+\gamma d k_{0||}[(\frac{19T_0'}{2p_0}-\frac{2T_0p_0'}{p_0^2}-\frac{\rho_0'}{\rho_0^2})(\gamma\eta_0''\epsilon\frac{4}{3}\nu - N g'\rho_0- 2N g\rho_0'-\gamma\rho_0\nu^2 )]\nonumber\\
-\gamma d
k_{0||}[(\frac{7T_0'}{p_0}-\frac{2T_0p_0'}{p_0^2})(\frac{4\gamma\eta_0'\epsilon\nu}{3}
- N g\rho_0 )
+(\frac{4\gamma\eta_0''\epsilon\nu}{3\rho_0} - N g'-\frac{ 2N g\rho_0'}{\rho_0}-\gamma\nu^2 )]\frac{f(z)'}{(f(z)-\nu)}\\
-\gamma d k_{0||}[(\frac{2T_0\rho_0''}{\rho_0}
 +\frac{35T_0'^2}{4T_0}
+ \frac{7T_0''}{2} +\frac{7T_0'\rho_0'}{\rho_0} -\frac{2T_0\rho_0'^2}{\rho_0^2})\frac{f(z)'}{(f(z)-\nu)}]\\
+ 3r\gamma [(3\rho_0'\rho_0 T_0^{-1}- \rho_0^2 T_0'T_0^{-2})-
\rho_0^2 T_0^{-1}\frac{f(z)'}{(f(z)-\nu)}]
 \nonumber\\
- (1+\gamma)p_0'\nu
 +\gamma p_0\nu\frac{f(z)'}{(f(z)-\nu)}
 +(f(z)-\nu)(\frac{4\gamma\eta_0'\epsilon}{3}\nu - N g\rho_0 )\\
{\cal E}(r,\epsilon,d,z)= -\gamma d k_{0||}[
(\frac{19T_0'}{2p_0}-\frac{2T_0p_0'}{p_0^2}-\frac{\rho_0'}{\rho_0^2})(N g\rho_0''+N g'\rho_0'+\gamma\rho_0'\nu^2)\\
-\gamma d k_{0||}[\frac{35T_0'^2}{2T_0p_0}
+\frac{7T_0''}{p_0}-\frac{14T_0'p_0'}{p_0^2}-\frac{2T_0p_0''}{p_0^2}+\frac{4T_0p_0'^2}{p_0^3}][N g\rho_0'+\gamma\rho_0\nu^2]\\
+\gamma d
k_{0||}[-\frac{63T_0'\rho_0'\rho_0''}{2\rho_0^2}+\frac{21T_0'\rho_0'^3}{\rho_0^3}+\frac{
105T_0'^3\rho_0'}{8\rho_0T_0^2}
+\frac{12T_0\rho_0'^2\rho_0''}{\rho_0^3}-\frac{6T_0\rho_0'^4}{\rho_0^4}-\frac{105T_0'^2\rho_0'^2}{4\rho_0^2T_0}+\frac{105T_0'^2\rho_0''}{4\rho_0T_0}\\
 -\frac{3T_0\rho_0''^2}{\rho_0^2}
 -\frac{4T_0\rho_0'\rho_0'''}{\rho_0^2}+\frac{21T_0'\rho_0'''}{2\rho_0} +\frac{\rho_0''''T_0}{\rho_0}
 - \frac{21\rho_0'T_0''\rho_0'}{2\rho_0^2}
+\frac{105T_0'T_0''\rho_0'}{4T_0\rho_0} +
\frac{7T_0'''\rho_0'}{2\rho_0}
 + \frac{21T_0''\rho_0''}{2\rho_0}]
\\
- \gamma d k_{0||}(\frac{N g\rho_0'''}{\rho_0}+\frac{N g''\rho_0'}{\rho_0}+\frac{2N g'\rho_0''}{\rho_0}+\frac{\gamma\rho_0''}{\rho_0}\nu^2)\\
 +\gamma d k_{0||}[\frac{N g\rho_0''}{\rho_0}+\frac{N g'\rho_0'}{\rho_0}+\frac{\gamma\rho_0'}{\rho_0}\nu^2+(\frac{7T_0'}{p_0}-\frac{2T_0p_0'}{p_0^2})(N g\rho_0'+\gamma\rho_0\nu^2)]\frac{f(z)'}{(f(z)-\nu)}\\
+\gamma d
k_{0||}[-\frac{T_0\rho_0'''}{\rho_0}-\frac{7T_0'\rho_0''}{\rho_0}+\frac{3T_0\rho_0'\rho_0''}{\rho_0^2}-\frac{2T_0\rho_0'^3}{\rho_0^3}-
\frac{35T_0'^2\rho_0'}{4p_0}
+\frac{7T_0'\rho_0'^2}{\rho_0^2}]\frac{f(z)'}{(f(z)-\nu)}\\
+3r\gamma [(\rho_0'^2T_0^{-1}-\rho_0 \rho_0'T_0'T_0^{-2}+ \rho_0
T_0^{-1}\rho_0'')
- \frac{T_0^{-1}\rho_0 \rho_0'f(z)'}{(f(z)-\nu)}] \\
-\nu p_0'' +p_0'\nu\frac{f(z)'}{(f(z)-\nu)}
 -(f(z)-\nu)(N g\rho_0'+\gamma\rho_0\nu^2)
 $\\
Here all quantities are dimensionless. The prime indicates
derivatives with respect to $z$.

\clearpage

\begin{figure} 
\centerline{\includegraphics[width=1\textwidth,clip=]{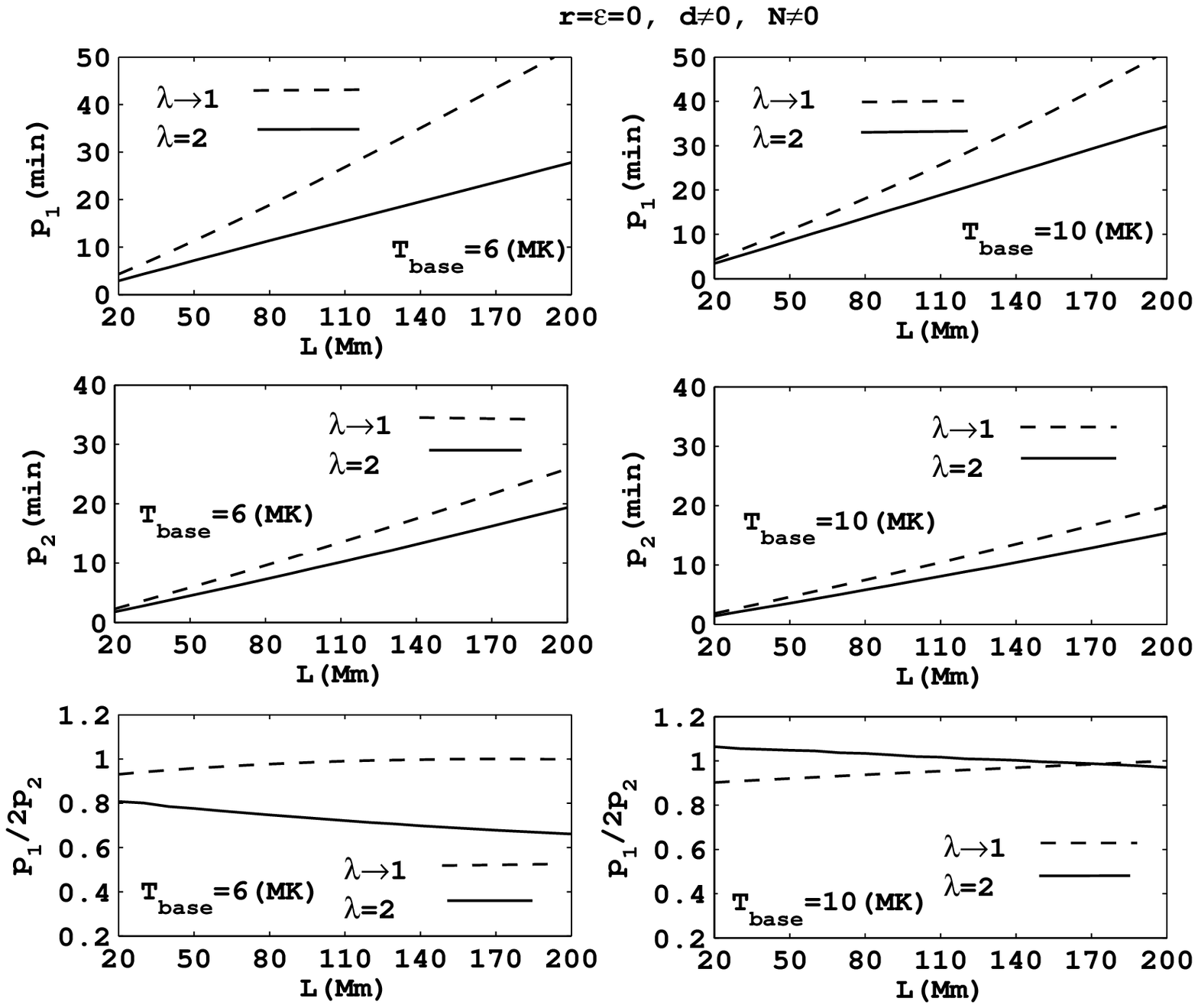}}
\caption[]{%
The periods of the fundamental mode $p_1$ (top row) and of the
first overtone $p_2$(middle row), and their ratios (bottom row)
for case 4 (see text) are plotted versus $L$ in oscillating loops
in the presence of conduction and gravity for different value of
$\lambda$ and $T_{\rm base}$. } \label{fig6}
\end{figure}

\begin{figure} 
\centerline{\includegraphics[width=1\textwidth,clip=]{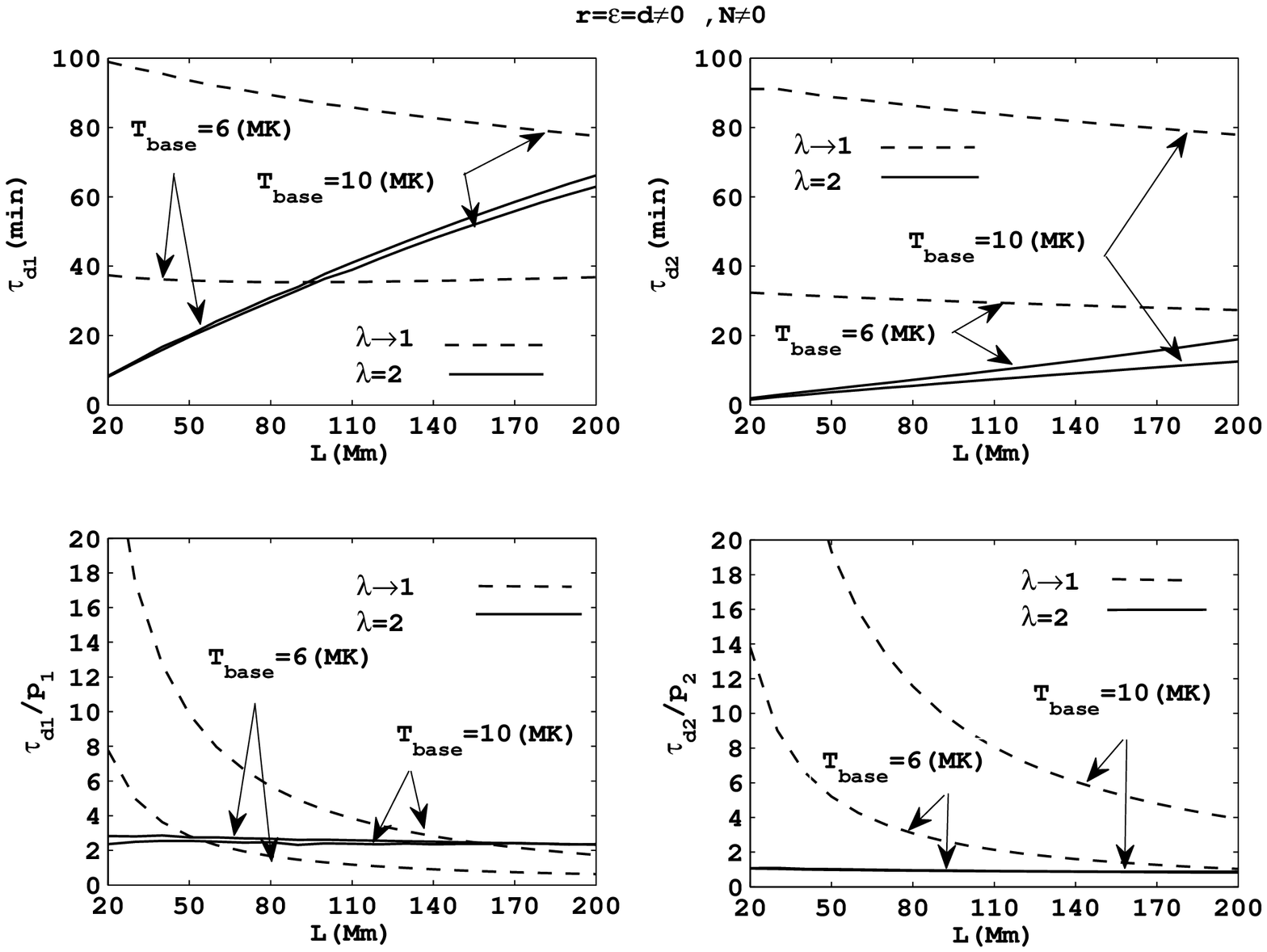}}
\caption[]{%
The damping times of the fundamental mode ($\tau_{\rm d1}$) and
of the first overtone ($\tau_{\rm d2}$) (top row), and the damping
times per period ($\tau_{\rm d1}/p_1, \tau_{\rm d2}/p_2$) (bottom
row) for case 4 (see text) are plotted versus $L$ in oscillating
loops in the presence of thermal conduction and gravity for
different value of $\lambda$ and $T_{\rm base}$. } \label{fig7}
\end{figure}

\begin{figure} 
\centerline{\includegraphics[width=1\textwidth,clip=]{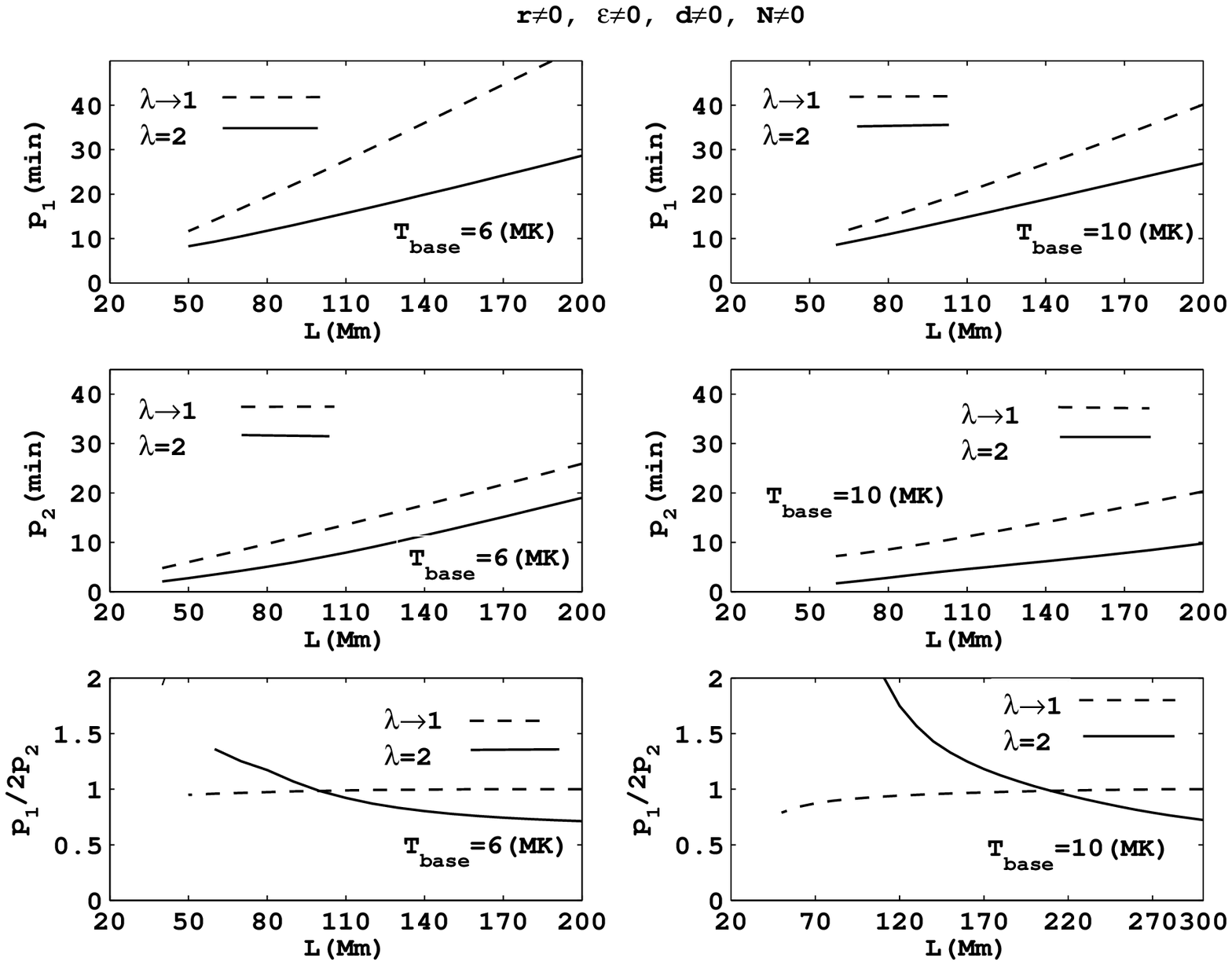}}
\caption[]{%
The periods of the fundamental mode $p_1$ (top row) and of the
first overtone $p_2$(middle row), and their ratios (bottom row)
for case 5 (see text) are plotted versus $L$ in oscillating loops
in the presence of radiation, viscosity, thermal conduction, and
gravity for different value of $\lambda$ and $T_{\rm base}$. }
\label{fig8}
\end{figure}

\begin{figure} 
\centerline{\includegraphics[width=1\textwidth,clip=]{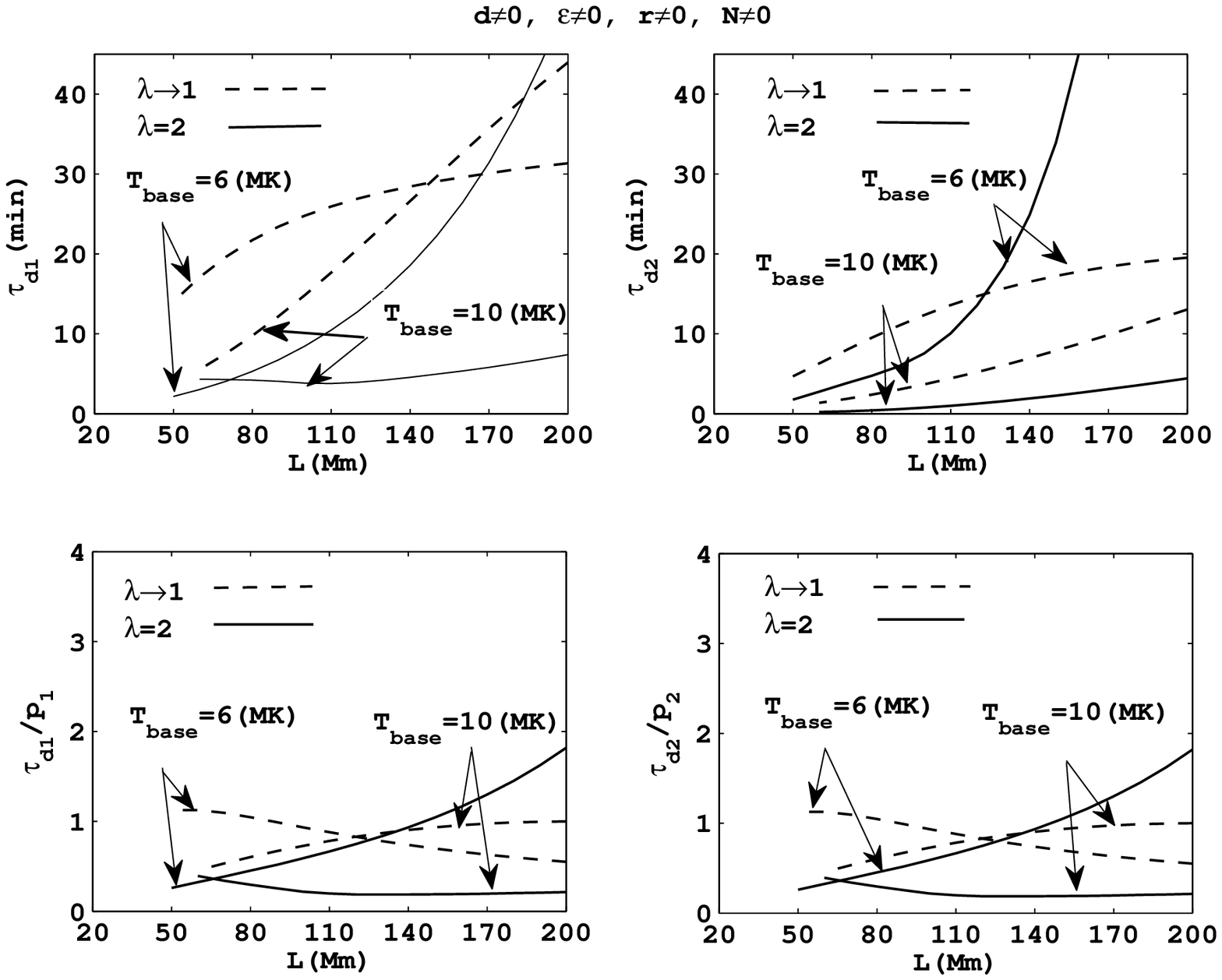}}
\caption[]{%
The damping times of the fundamental mode ($\tau_{\rm d1}$) and of
the first overtone $(\tau_{\rm d2}$) (top row), and the damping
times per period ($\tau_{\rm d1}/p_1$, $\tau_{\rm d2}/p_2$)
(bottom row) for case 5 (see text) are plotted versus $L$ in
oscillating loops in the presence of radiation, viscosity, thermal
conduction, and gravity for different value of $\lambda$ and
$T_{\rm base}$. } \label{fig9}
\end{figure}

\end{article}

\end{document}